\documentstyle[12pt]{article}
\begin{document}
\def\Tm{T^m}
\def\tm{t_m}
\def\Tdm{T^m_d}
\def\tdm{t_m^d}
\def\ot{\otimes}
\def\br{\bf R}
\def\al{\alpha}
\def\bt{\beta}
\def\th{\theta}
\def\ga{\gamma}
\def\vth{\vartheta}
\def\de{\delta}
\def\lm{\lambda}
\def\b{\beta}
\def\k{\kappa}
\def\om{\omega}
\def\si{\sigma}
\def\w{\wedge}
\def\od{\sqrt{L^2+1}}
\def\Tn{T^n}
\def\tn{t_n}
\def\Tdn{T^n_d}
\def\tdn{t_n^d}
\def\e{\varepsilon}
\def\ti{\tilde}
\def\js{{1\over 4}}
\def\I{{\cal I}}
\def\L{{\cal L}}
\def\D{{\cal D}}
\def\G{{\cal G}}
\def\bz{{\bar z}}
\def\E{{\cal E}}
\def\B{{\cal B}}
\def\M{{\cal M}}
\def\K{{\cal K}}
\def\J{{\cal J}}
\def\tJ{\ti{\cal J}}
\def\R{{\cal R}}
\def\d{\partial}
\def\la{\langle}
\def\ra{\rangle}
\def\bc{{\bf C}}
\def\st{\stackrel{\w}{,}}
\def\lta{\leftarrow}
\def\rta{\rightarrow}
\def\scu{$SL(2,\bc)/SU(2)$~ $WZW$  }

\def\be{\begin{equation}}
\def\ee{\end{equation}}
\def\jp{{1\over 2}}

\def\tD{\Delta^*}
\def\slc{SL(2,{\bf C})}
\begin{titlepage}
\begin{flushright}
{}~
IML 02-XY\\
hep-th/0210095
\end{flushright}

\vspace{1cm}
\begin{center}
{\Huge \bf Yang-Baxter $\si$-models
and dS/AdS T-duality}\\
[50pt]{\small
{\bf Ctirad Klim\v{c}\'{\i}k}
\\ ~~\\Institute de math\'ematiques de Luminy,
 \\163, Avenue de Luminy, 13288 Marseille, France}
\end{center}

\vspace{0.5 cm}
\centerline{\bf Abstract}
\vspace{0.5 cm}

 We point out the existence of  nonlinear $\si$-models  on group
manifolds
 which are left symmetric and right
    Poisson-Lie symmetric.    We discuss  the corresponding rich
T-duality story with particular emphasis on two ${\rm examples : }$
the
 anisotropic principal chiral model and the
 $\scu$ model. The latter has the de Sitter space as its (conformal)
non-Abelian  dual.

\end{titlepage}
\section{Introduction}

This paper is devoted to the study of a particular
class of $\si$-models living on group manifolds. The metric and the
three-form $H$
of such a model are {\it left}-invariant, however,  the model is
 Poisson-Lie symmetric with respect to the {\it right} action of the
group on itself.
We shall refer to such $\si$-models as to the Yang-Baxter $\si$-models
 for reasons
which will become clear in what follows.

The Yang-Baxter $\si$-models have interesting algebraic properties. First
of all, they have a very rich T-duality story because they can be dualized
either
from the left (by the traditional non-Abelian T-duality \cite{OQ}) or from
the right (by the Poisson-Lie T-duality \cite{KS1}). They have also another
 remarkable
feature: the phase space of a Yang-Baxter $\si$-model  can be
 represented by two Drinfeld doubles  {\it nonisomorphic}
as {\it groups}. This leads e.g. to some
insights about the structure of the   $\scu$ theory.

In what follows,  we  describe the general formalism and then
 study the Yang-Baxter $\si$-models on   a simple compact
Lie group $K$ and on the group $AN$ in the Iwasawa decomposition
of   the complexified group $K^\bc=KAN$. For $K^{\bf C}=SL(2,\bc)$,
the former
 example turns out to be the anisotropic
principal chiral model (known to be integrable \cite{Ch}) and
the latter
the (conformal) $SL(2,\bc)$/$SU(2)$
WZW model describing the strings on the Euclidean $AdS_3$
target.  The $AdS_3$ model turns out to have the  de Sitter background
 $dS_3$
(with an appropriate three-form field) as its Poisson-Lie T-dual.

In section 2, we review the notion of the (right) Poisson-Lie symmetry of
nonlinear $\si$-models and show how the model enjoying this property
can be rewritten in a  first order form by using the concept of the
Drinfeld
double.
Then we show that the additional left symmetry
generically
means that the geometry of the target is given by the elements
of the $r$-matrix  defining the Poisson-Lie structure on the PL symmetry
group. This
gives the name to our models. We finish the section by considering
the Poisson-Lie symmetry  of (non-unitary) $\si$-models
having a {\it real}
Euclidean action.

In section 3, we explicitely describe the Poisson-Lie symmetry
of our two principal examples of the Yang-Baxter $\si$-models: the
 anisotropic
principal chiral  model on compact groups $K$ and the $K^{\bf C}/K$
WZW theory. In section 4, we explain why the Poisson-Lie symmetry
generically means the T-dualizability of the model and
 we describe the T-duals of the anisotropic model on $SU(2)$
and of the $\scu$ model.  Then  we show how the rich
Kac-Moody symmetry of the $\scu$ model translates under the duality
to the same symmetry of the de Sitter model. We finish by short
conclusions.

\vskip1pc

\section{Poisson-Lie symmetry}
In this section, we give a method for the  construction
of the Yang-Baxter
$\si$-models.
  \vskip1pc
\noindent 1. As it is well-known, a two-dimensional non-linear
$\si$-model
 is
a field theory canonically associated  to  a metric ( symmetric tensor)
$G_{ij}$
and a two-form (antisymmetric) tensor $B_{ij}$ on some manifold $M$.
Its action is given (in some local coordinates $x^i$) as
$$S=\jp\int d\si d\tau  (G_{ij}(x) +B_{ij}(x))\d_+x^i\d_- x^j\equiv
\jp\int
 d\si d\tau E_{ij}(x)\d_+x^i\d_- x^j,\eqno(1)$$
where
$$\d_\pm=\d_\tau\pm \d_\si, \quad \xi^\pm=\jp(\tau\pm\si)$$
and $\tau$ and $\si$ are respectively the worldsheet time and
 space coordinates. Suppose that there is a free right action
of a Lie group $G$  on $M$ generated by vector fields $v_a(x)$ on $M$
and consider  the variation of the action (1) with respect to the
$G$-transformations
with the world-sheet dependent parameters $\e^a(\si,\tau)$:
$$\delta S\equiv S(x+\e^av_a)-S(x)=\jp\int d\si d\tau
 \e^a(\L_{v_a}E_{ij})\d_+x^i\d_- x^j
+\int J_a\w d\e^a.$$
Here  $\L_{v_a}$ means the Lie derivative with respect to $v_a$ and
the worldsheet current one-forms $J_a$ are
$$J_a=- v_a^i(x)E_{ij}(x)\d_-x^j d\xi^-+  v_a^i(x)E_{ji}(x)\d_+x^j
d\xi^+.\eqno(2)$$
We say (following \cite{KS1}) that the $\si$-model (1) is
$\ti G$-Poisson-Lie symmetric
if
$$\delta S=\int \e^a(dJ^a-\jp \ti f^{~kl}_{a}J_k\w J_l),\eqno(3)$$
where $\ti f^{~kl}_{a}$ are the structure constantes of
$\ti\G\equiv Lie(\ti G)$.
  The Poisson-Lie
symmetry condition can be written also in terms of the $\si$-model
tensor $E_{ij}$
as follows
$$\L_{v_a}E_{ij}=-\ti f^{~kl}_{a}v^m_k v^n_l E_{mj}E_{in}.\eqno(4)$$

\vskip1pc
\noindent 2.
If the manifold $M$ is itself the group $G$, there is a powerful
 method \cite{KS1}
of finding the
right Poisson-Lie symmetric $\si$-models (i.e. of solving
the condition (4)). It uses the crucial concept of the
Drinfeld double\footnote{The Drinfeld double  $D$ is a  $2n$-dimensional
 Lie group whose
Lie algebra $\D =Lie(D)$, viewed  as the linear space, can be decomposed
into a direct sum of vector spaces $\G$ and $\ti\G$,  which are themselves
 maximally isotropic subalgebras of $ D$ with respect
to a non-degenerate symmetric invariant bilinear form $(.,.)_\D$
on $\D$ \cite{D,FG}.
The Lie groups corresponding to the Lie algebras $\G$ and $\ti\G$ we denote
as $G$ and $\ti G$ and we shall suppose,   that
each element $l$ of $D$ can be decomposed in unique way as the product
$l=gb$, where $g\in G$ and $b\in \ti G$. Moreover, the induced
map $D\to G\times \ti G$ should be a diffeomorphism. In the  same way
we suppose that there is a dual global smooth and
unambiguous decomposition $D=\ti GG$.} of $G$ and $\ti G$
and constructs the $\si$-model (1)
by starting from its first order Hamiltonian action written on the phase
 space.
Recall that (for any dynamical system) the latter has always the form
$$S=\int (\theta -Hd\tau),\eqno(5)$$
where $d\theta$ is the symplectic form and $H$ a Hamiltonian fonction.

The crucial statement \cite{KS1} is as follows:
 the  phase space of  right $\ti G$-Poisson-Lie symmetric $\si$-model
 on a Lie group
target $G$    can be identified with the
loop group $LD$ of the Drinfeld double $D$ \cite{KS2}. Thus a trajectory
in the phase space is a map $l(\si,\tau)$ from the worldsheet into $D$
where $\tau$ is interpreted as the evolution parameter.  The
 first order action (5) of the  right $\ti G$-Poisson-Lie symmetric
$\si$-model
 on a Lie group
target $G$ then reads
$$S=\jp\int d\si d\tau (\d_\tau ll^{-1},\d_\si ll^{-1})_\D
+{1\over 12}\int d^{-1}(dll^{-1}\st [dll^{-1}\st dll^{-1}])_\D$$
$$-\jp\int d\si d\tau (\d_\si ll^{-1},\E\d_\si ll^{-1})_\D.\eqno(6)$$
 The (chiral) WZW action in the first line
correspond the term $\theta$ in (5) and the Hamiltonian term is on the
second line.
It involves the operator $\E:\D\to\D$ fulfilling
\vskip1pc

\noindent 1) $(\E x,y)_\D=(x,\E y)_\D$ ($\E$ is a selfadjoint);

\noindent 2) $\E^2=Id$ ($\E$ squares to the identity operator);

\noindent 3) The   $G$-dependent operator $\J\E_g\ti\J:\ti\G\to\G$
is invertible.
\vskip1pc
\noindent  Here   $\J(\tJ):\D\to\D$ is the  projector  to  $\G$ ($\ti\G$)
with the
kernel $\ti\G$ ($\G$) and $\E_g\equiv Ad_{g^{-1}}\E Ad_g$. It is important
to stress that $Ad_g$ means the adjoint action of an element
$g\in G(\subset D)$
on $\D$.
  \vskip1pc
\noindent  3. Let us sketch the derivation of the second order Lagrangian
 of the type (1)
from the first order action (6).
Our assumptions
about the structure of the Drinfeld double $D$ allow to
decompose each trajectory $l(\si,\tau)$ as
$$l(\si,\tau)=g(\si,\tau)\ti h(\si,\tau),\qquad g\in G,
\quad \ti h\in \ti G.\eqno(7)$$
Inserting this decomposition into the action (6) and using the standard
Polyakov-Wiegmann formula ($g^*,\ti h^*$ mean the pull-backs of the maps
$g,\ti h:D\to D$)
$$(g\ti h)^*c=g^*c +\ti h^*c -d(g^*(l^{-1}dl)\st \ti h^*(dll^{-1}))_\D;$$
$$c={1\over 6}(dll^{-1}\st [dll^{-1}\st dll^{-1}])_\D,$$
 we obtain
$$S=\int (\d_\si\ti h\ti h^{-1},g^{-1}\d_\tau g)_\D-
\jp\int(g^{-1}\d_\si g +
\d_\si\ti h\ti h^{-1},\E_g(g^{-1}\d_\si g +
\d_\si\ti h\ti h^{-1}))_\D,\eqno(8)$$
where  and we tacitly suppose the measure
$d\si d\tau$ present in the formula. Now we note that the  expression (8)
is Gaussian in the $Lie(\ti G)$-valued variable
$\d_\si\ti h\ti h^{-1}$ which permits us to solve it away and to obtain
a $\si$-model
$$S=\jp\int d\si d\tau (g^{-1}\d_+ g,
(\J \E_g\ti\J)^{-1}(\J+\J \E_g\J)g^{-1}\d_- g)_\D.
\eqno(9)$$
In this derivation we have used the self-adjointness
of $\E$ and the following consequences of the identity
$\E^2=\E_g^2=Id$ :
$$(\J \E_g\ti\J)^{-1}=\ti\J \E_g\J-\ti\J \E_g\ti\J
(\J \E_g\ti\J)^{-1}\J \E_g\J;$$
$$(\J \E_g\ti\J)^{-1}\J \E_g\J=-\ti\J \E_g\ti\J (\J \E_g\ti\J)^{-1}.$$
We invite the reader to check, that, indeed, the resulting $\si$-model (9)
 is right $\ti G$-Poisson-Lie symmetric.  It is also immediate that the
left invariance
of (9)
takes place if $\E_g=\E$, i.e. if $\E$ is a $G$-invariant operator on $\D$.
\vskip1pc
\noindent {\bf Conclusion}: let $D$ be  the Drinfeld double of the
 group $G$ and
$\E$ the self-adjoint,
unipotent,
$G$-invariant operator on $\D$, such that $\J\E\ti\J$ is invertible.
 Then the
action
$$S=
\jp\int d\si d\tau (g^{-1}\d_+ g,\ti Rg^{-1}\d_- g)_\D,\eqno(10)$$
with $$\ti R\equiv  (\J \E\ti\J)^{-1}(\J+\J \E\J) $$
defines the Yang-Baxter $\si$-model on $G$.
\vskip1pc

\noindent 4. The reader might wish to see a direct demonstration that
the Yang-Baxter
$\si$-model (10) is right Poisson-Lie symmetric. For this, let us first
calculate the
current forms $J_a$ according to (2):
$$J_aT^a=- \ti Rg^{-1}\d_-g d\xi^- + \ti R^\dagger g^{-1}\d_+g d\xi^+,$$
where  $T^a$ is the basis of $\ti\G$ in which the structure constants are
$\ti f^{~kl}_{a}$.
It is easy to see that the Hamiltonian field equations derived from the
 first order
action (6) read
$$(Id\mp\E)\d_\pm ll^{-1}=0.$$
Using the decomposition $l=g\ti h$, the $G$-invariance of $\E$ an
the the fact that  $\ti R^\dagger=(\J \E\ti\J)^{-1}(\J-\J \E\J)$, we
 obtain
 $$\d_-\ti h\ti h^{-1}=-\ti R g^{-1}\d_-g,\quad \d_+\ti h\ti h^{-1}=
\ti R^\dagger g^{-1}\d_+g\eqno(11)$$
and from (11) it clearly follows that the field equations of the
Yang-Baxter
$\si$-model have the zero-curvature form (3):
$$dJ_a= \jp \ti f^{~kl}_{a}J_k\w J_l.$$

\vskip1pc
\noindent 5.
Now we turn to the explication of the title of this article.
Suppose that the operator $\ti R:\G\to\ti\G$ has an inverse
$R:\ti\G\to\G$.
The properties of $\E$ then imply the  following identity
$$ \J Ad_g \tJ Ad_{g^{-1}}\tJ=\J Ad_g\J R\ti\J Ad_{g^{-1}}\ti\J -R.
\eqno(12)$$
The $g$-dependent operator from $\ti\G$ into $\G$ on the l.h.s. of (12)
deserves a special name;
we call it   $\Pi_g$:
$$\Pi_g=\J Ad_g \tJ Ad_{g^{-1}}\tJ.\eqno(13)$$
In fact, $\Pi_g$ turns out to be the Poisson-Lie structure on $G$
induced by the double $D$ \cite{KS2}.  For completeness,
we write the corresponding Poisson-Lie bracket of two functions
$\phi,\psi$
on $G$:
$$\{\phi,\psi\}_G=(\nabla_L\phi,\Pi_g\nabla_L\psi)_\D.\eqno(14)$$
Here the differential operator $\nabla_L$ is considered to live in
$\ti\G$.
It is defined by
$$(\nabla_L,x)_\D=\nabla_L^x,$$
where   $\nabla_L^x$ is the differential operator acting
on functions on $G$ and corresponding to the left infinitesimal action
of the Lie algebra element $x\in\G$.

Now note that the relations (12,13)   mean
that  the Poisson-Lie
bracket (14) on $G$ can be rewritten in the following form
$$\{\phi,\psi\}_G=(\nabla_R\phi,R\nabla_R\psi)_\D
-(\nabla_L\phi,R\nabla_L\psi)_\D,$$
where $\nabla_{R(L)}$ corresponds to the right (left)  infinitesimal
 action on
$G$. It is now clear that $R$ gives rise to
 an element $r\in\G\otimes\G$
such that the Poisson-Lie  bracket can be rewritten in the
quasitriangular
way:
$$\{\phi,\psi\}_G=\la r, \nabla_R\phi\ot \nabla_R\psi\ra
-\la r, \nabla_L\phi\ot \nabla_L\psi\ra . $$
Here $\la .,.\ra$ denotes the canonical pairing between the
vector space  and its dual.
Since the Poisson bracket is antisymmetric, we see that the symmetric part
$r_S$ of $r$ is $G$-invariant. The  Jacobi identity, in turn,
then implies the (modified ) Yang-Baxter equations for the antisymmetric
part $r_A$ of $r$:
$$Ad_G[r_A,r_A]_{Sch}=0,$$
where $[.,.]_{Sch}$ is the algebraic Schouten bracket.

Thus we observe that  the choice of the $G$-invariant selfadjoint
unipotent
$\E$ with invertible $\ti R$
  gives rise to the $r$-matrix
encoding the (quasitriangular) Poisson-Lie structure on $G$. This
explains
the adjective "Yang-Baxter" which we have attributed to the left-invariant
and right Poisson-Lie symmetric $\si$-models (10). By a slight abuse
of the terminology, we stick to this name also in  cases where $\ti R$
is not invertible.
\vskip1pc
\noindent 6. It is well-known that the unitary $\si$-models do not have
a real action on the Euclidean world-sheet. Nevertheless, there
is a class of important (though non-unitary) Euclidean models which
do have  the real
action. The most prominent example
of such a theory is the $\scu$ model describing strings on the
Euclidean $AdS_3$. Here we define the notion of the Poisson-Lie symmetry
for
the  class of the Euclidean $\si$-models
  with the real action
$$S=\jp\int d\si d\tau  E_{ij}(x)\d_z x^i\d_\bz x^j,$$
where
$$\d_z=\d_\tau+i\d_\si,\quad \d_\bz=\d_\tau-i\d_\si.$$
The reality of the Lagrangian require the matrix $E_{ij}$ to be Hermitean
($E_{ij}=\bar E_{ji}$). The definition (2) of the current one-forms $J_a$
 now gives
$$J_a= i v_a^i(x)E_{ji}(x)\d_z x^j dz- i v_a^i(x)E_{ij}(x)\d_\bz x^j
 d\bar z,$$
where $$z=\jp(\tau-i\si),\quad \bar z=\jp(\tau+i\si).$$
With this notation, the condition of the right Euclidean Poisson-Lie
symmetry is given by the
same expression (3)  as in the Minkowski case. The Euclidean analogue
of (4) reads
$$\L_{v_a}E_{ij}=-i\ti f^{~kl}_{a}v^m_k v^n_l E_{mj}E_{in}.\eqno(15)$$
 In the Euclidean case, it is also easy to generate right Poisson-Lie
symmetric models
from the action (6) on the double. The only modification is   that now
 $\E$
in (6) has to fulfil  $\E^2=-Id$.
In particular, the resulting
Yang-Baxter $\si$-model on the Euclidean worldsheet has the Lagrangian
$$L=
\jp\int d\si d\tau (g^{-1}\d_z g,\ti R_{eu}g^{-1}\d_\bz g)_\D\eqno(16)$$
 with $$\ti R_{eu}=(\J\E\ti\J)^{-1}(\J-i\J\E\J).$$

\section{Examples}
\subsection{Anisotropic principal chiral model}
\noindent 1. Now it is time to give  examples.  For the group $G$ we
take a simple
connected and simply connected compact group $K$ and for its Drinfeld
double $D$ its complexification $D=K^{\bf C}$. So, for instance, the
double of $SU(2)$ is $SL(2,\bf C)$. The invariant bilinear form on
$\D=Lie(D)$ is given by
$$(x,y)_\D={\rm Im}T(x,y),$$
or, in other words, it is just the imaginary part of the Killing-Cartan
form $T(.,.)$ normalized in the way that the square of the
lenght of the longest root is equal to two.
 Since  $G$ is the real form of $K^{\bc}$, clearly,
the imaginary part of $T(x,y)$ vanishes if $x,y\in\G$. Hence, $G=K$ is
isotropically embedded in $K^{\bc}$.

The dual group $\ti G$ coincides with
the so called $AN$ group in the Iwasawa decomposition of $K^\bc$:
$$  K^\bc= KAN.\eqno(17)$$
For the groups $SL(n,{\bf C})$ the group $AN$ can be identified with
upper triangular matrices of determinant $1$
and with positive real numbers on the diagonal.
In general, the elements of $AN$ can be uniquely represented by means
of the exponential map as follows
$$ b={\rm e}^{\phi}{\rm exp}[\Sigma_{\al>0}v_\al E_\al]\equiv
 {\rm e}^{\phi}n.\eqno(18)$$
Here $\al$'s denote the roots of $\K^\bc$, $v_\al$ are complex numbers
and $\phi$ is an Hermitian element\footnote{Recall that the
Hermitian element of  any complex simple Lie algebra $\K^\bc$
is an eigenvector of the
involution which defines the compact real form $\K$;
the corresponding eigenvalue
is $(-1)$ . This involution comes from the group involution
$g\to (g^{-1})^\dagger$.
The anti-Hermitian elements that span
 the compact real form are eigenvectors
of the same involution with the eigenvalue equal to $1$. For elements
of $sl(n,{\rm C})$ Lie algebra, the Hermitian element is
indeed a Hermitian
matrix in the standard sense of this term.}
 of the Cartan subalgebra
of $\K^\bc$. Loosely said, $A$ is the "noncompact part" of the
complex maximal
torus of $K^\bc$. The isotropy of the Lie algebra $\ti\G$ of  $\ti G=AN$
follows from (18); the fact that $\G$ and $\ti \G$ generate together
the Lie algebra $\D$ of the whole double  is evident from (17).
\vskip1pc
\noindent 2.
We wish to construct (Minkowskian) Yang-Baxter  $\si$-models on the
compact target
$K$. Consider the following $\br$-linear operator $\E_1:\D\to \D$:
$$\E_1x=   \I x^{\dagger}.\eqno(19)$$
Here $\I$ it is the multiplication by the
imaginary unit on $\K^\bc$(
  viewed as the  {\it real} Lie algebra).
It is easy to see that $\E_1^2=Id$ and $\E_1$ is selfadjoint,
i.e. for any $x,y\in\D$ it holds $(\E_1 x,y)_\D=(x,\E_1 y)_\D$.
Moreover, $\E_1$ is $K$-invariant as
$$\E_1 Ad_k x=\E_1 (kxk^\dagger)=\I kx^\dagger k^\dagger=
Ad_k\E_1 x.$$
Thus, $P_1=\jp(1+\E_1)$ is the $K$-invariant  idempotent. In fact,
 it belongs
to a one-parametric family of the invariant  idempotents denoted
$P_\al$ who project on the subspaces
$(\al Id-\I)\K\subset\K^{\bc},\al> 0$
and whose   kernels are $(\al Id+\I)\K$. Since the kernel is clearly
orthogonal to the image of the projector, the idempotent $P_\al$
 is selfadjoint.
Moreover  $P_\al$ is $K$-invariant, since both its image and kernel
are $K$-invariant. Finally, it can be easily verified that
$\J (2P_\al-Id)\ti\J$ is invertible. Thus the choice $\E_\al=2P_\al-Id$
gives a one parameter family of the Yang-Baxter $\si$-model on $G$
 with the Lagrangian given by (10).

\vskip1pc
\noindent 3.
The reader may wish to visualize more explicitely
 what the Yang-Baxter $\si$-model is about for the simplest
case of $K=SU(2)$. For this, it is convenient to choose a basis
$T^i=\jp i\si^i, i=1,2,3$
in $su(2)\equiv Lie (SU(2))$,
where $\si^i$ are the standard Pauli matrices:
$$T^1=\jp\left(\matrix{0&i\cr i&0}\right),\quad
 T^2=\jp\left(\matrix{0&1\cr -1&0}\right), \quad
T^3=\jp\left(\matrix{i&0\cr 0&-i}\right).\eqno(20a)$$
The dual basis in $Lie(AN)$ is then
$$t_1=\left(\matrix{0&2\cr 0&0}\right),\quad
 t_2=\left(\matrix{0&-2i\cr 0&0}\right), \quad
t_3=\left(\matrix{1&0\cr 0&-1}\right).\eqno(20b)$$
Here by the duality we mean the following relation between the
 two basis:
$$(T^i,t_j)_\D=\delta^{i}_{~j}.$$
It is not difficult to evaluate the operator
 $\ti R_\al=( \J \E_\al\ti\J)^{-1}( \J +\J\E_\al\J)$, where $\E_\al=
2P_\al-1$. It is given by
$$\ti R_\al T^1= {(\al t_1- t_2)\over 2(\al^2+1)} ,
\quad
\ti R_\al T^1= {(\al t_2+ t_1)\over 2(\al^2+1)} ,\quad
 \ti R_\al T^3= {1\over 2\al}t_3.\eqno(21)$$
The Lagrangian (xy) of the Yang-Baxter
 $\si$-model on $SU(2)$ can be now
written as
$$L={1\over 4\al}(k^{-1}\d_+k)_3(k^{-1}\d_-k)_3+$$ $$+
{\al\over 4(\al^2+1)} (k^{-1}\d_+k)_1(k^{-1}\d_-k)_1
+{\al\over 4(\al^2+1)}(k^{-1}\d_+k)_2
(k^{-1}\d_-k)_2 .\eqno(22)$$
Here we have  decomposed $k^{-1}dk$ as  $(k^{-1}dk)_jT^j$ and neglected
the
total derivatives. By the way, the model (22) is known to be
integrable and it is usually referred to as the anisotropic
principal chiral model. It is interesting to note that we have discovered
a new algebraic structure of this model which will probably contribute
to a better understanding of its quantum properties.

It is also important to remark, that the obtained model (22) is singular
for $\al=0$ and $\al\to\infty$. Indeed, for these
values of the parametres the idempotents $P_\al$ are
not well defined  (the kernel would be the same as the image!)
 The  singularities can be very naturally
interpreted also from the geometrical point of view.
 Indeed, for $\al\to 0$, the ratio of the  (anisotropic)
third coefficient with respect to the first (or second) one goes to
infinity.
This is the extreme anisotropic limit where the $\si$-model geometry
degenerates in one direction. In the case $\al\to\infty$ this ratio
approaches $1$, which corresponds the standard (isotropic) principal
chiral model. However, the singularity shows up
in the overall coefficient going to zero.

\subsection{$K^{\bc}/K$ WZW model}

\noindent 1.
Our next  example has the real action in the Euclidean signature. Its
underlying
 Drinfeld double is the same group $K^\bc$ as in the previous one
but the role of the groups $G$ and $\ti G$ gets reversed. This means
that the $K^{\bc}/K$ WZW model lives on the group $G=AN$ and it is
right Poisson-Lie symmetric with respect to the dual group $\ti G=K$.

The first order action of the $K^{\bc}/K$ WZW model is given by the
expression (6) with $\E=\I$ (cf. (19)).  Clearly, $\I$ is selfadjoint
 (yes,
it sometimes happens that the multiplication by the imaginary unit
is a selfadjoint operator...), $\I$ is $K^\bc$-invariant, $\I^2=-Id$ and
$\J\I\ti\J$ is invertible. Thus   $K^{\bc}/K$ WZW theory is the
Yang-Baxter $\si$-model with   the second
order real Euclidean Lagrangian (16) and  $\ti R_{eu}=
(\J\I\ti\J)^{-1}(\J-i\J\I\J).$
\vskip1pc
\noindent 2.
As in the previous example, we give the detailed description of the
case $K^\bc=SL(2,\bc)$. We shall work with the same basis of $\D$ as
before but now we remember that the basis $t_a$ is that of $\G=Lie(AN)$
and
the dual one $T^a$ that of $\ti \G=\K$.
 The operator $\I$ reads
$$\I t_1=-t_2,\quad \I t_2=t_1,\quad \I t_3 =2 T^3;$$
$$\I T^1=-\jp t_1 +T^2,\quad \I T^2=-\jp t_2 -T^1,\quad \I T^3=
-\jp t_3.\eqno(23)$$
  It can be directly checked that
 $\I$ is selfadjoint with respect to the form $(.,.)_\D$,
it is $SL(2,\bc)$-invariant and
it  evidently holds  $\I^2=-Id$.
With the choice  $\E=\I$, the first order
  action (6) takes the  form:
$$S= \jp\int d\si d\tau {\rm ImTr}(\d_\tau ll^{-1}\d_\si ll^{-1})
+{1\over 12}\int d^{-1}{\rm ImTr}(dll^{-1}\w [dll^{-1}\st dll^{-1}])$$
$$-\jp\int d\si d\tau {\rm ReTr}(\d_\si ll^{-1}\d_\si ll^{-1})$$
or, even more simply,
$$S=\jp\int d\si d\tau {\rm ImTr}(\d_\bz ll^{-1}\d_\si ll^{-1})
+{1\over 12}\int d^{-1}{\rm ImTr}(dll^{-1}\w [dll^{-1}\st dll^{-1}]).
\eqno(24)$$
In   words: the  first order action
of the $\scu$ model can be written
as a sort of semi-chiral WZW model on the complex group $SL(2,\bc)$.
\vskip1pc
\noindent 3.
 The resulting second order Lagrangian on $G=AN$
can be obtained by decomposing
$$l=bk,\quad b\in AN,k\in SU(2),\eqno(25)$$
and eliminating $\d_\si kk^{-1}$. The result is (cf. (16))
 written as
 $$\ti L=\jp  (b^{-1}\d_z b, \ti R_{eu} b^{-1}\d_\bz  b)_\D,\eqno(26)$$
with $$\ti R_{eu}=
(\J \I\ti\J)^{-1}(\J-i\J\I\J).$$
From (20ab) and (23), it is easy to calculate the
 operator $\ti R_{eu}:\G\to\ti\G$:
$$\ti R_{eu}t_1= -2(T^1 +iT^2),\quad \ti R_{eu}t_2=-2(T^2-iT^1),
\quad \ti R_{eu}t_3=-2T^3.
\eqno(27)$$
 Using   the following parametrization of the group $AN$:
$$b=\left(\matrix{e^{\phi}&2e^{-\phi}(u_1-iu_2)\cr 0&e^{-\phi}}\right),
\eqno(28)$$
we  calculate
  $b^{-1}db =(b^{-1}db)^jt_j$:
$$(b^{-1}db)^1=e^{-2\phi}du_1,(b^{-1}db)^2=e^{-2\phi}du_2 ,\quad
\quad (b^{-1}db)^3=d\phi.\eqno(29)$$
Putting together (26), (27) and (29) and setting  $u=u_1+iu_2$,
 we obtain that the Yang-Baxter Lagrangian
(26) gives the action
  of the $\scu$ model (cf.\cite{Gaw,Ts}):
$$S=-\int d\si d\tau (\d_z\phi \d_\bz\phi+e^{-4\phi}\d_z u \d_\bz \bar u).
\eqno(30)$$
Up to an unimportant overall normalization,
the background metric and  $H$-field of the model (xy) are
$$ds^2_{AdS}=d\phi d\phi+e^{-4\phi}dud\bar u,\quad H=-2 e^{-4\phi}d\phi\w
du\w d\bar u.\eqno(31)$$
If we set $\ti R_{ab}=(t_a,\ti R_{eu} t_b)_\D$, the condition (15)
expressing
the right Poisson-Lie symmetry  of (26) can be rewritten as
$$f_{ca}^{~~n}\ti R_{nb}+f_{cb}^{~~n}\ti R_{an}=
i\ti f_c^{~kl}\ti R_{kb}\ti R_{al}.
\eqno(32)$$
Here $f_{ca}^{~~n}$ are the structure constants
   of $Lie (AN)$ in the basis (20b) and $\ti f_c^{~kl}$ of $su(2)$
in the basis (20a). It is easy to verify that our operator $\ti R_{eu}$
 given by (27)
satisfies (32).

The $su(2)$-valued Poisson-Lie current $J$ is given by
$$J=i\ti R^\dagger g^{-1}\d_z g dz -i\ti R  g^{-1}\d_\bz g d\bz=$$
$$=-2ie^{-2\phi}\d_z u(T^1-iT^2)dz  -2i\d_z\phi T^3 dz
+2ie^{-2\phi}\d_\bz \bar u(T^1+iT^2)dz  +2i\d_\bz\phi T^3 d\bz.$$
It is easy to verify that the zero curvature condition $dJ=J\w J$ is
 equivalent
to the field equations of the $\scu$ model:
$$\d_z\d_\bz \phi+2e^{-4\phi}\d_z u\d_\bz \bar u=0, $$
$$\d_z(e^{-4\phi}\d_\bz \bar u)=0,\quad \d_\bz(e^{-4\phi}\d_z  u)=0.$$
\vskip1pc
\noindent 4. {\bf Note}:
The Yang-Baxter property
of the  $\scu$-model leads naturally to two different representations
of the phase space of the theory: either
  as the complex loop group $LSL(2,\bc)$
or as the loop group of the cotangent bundle of $AN$. The latter
representation
(which is naturally induced by the left invariance)
is well-known and discussed
  in literature (cf. \cite{Ts}). Here we advance the former one.
Its advantage may consist
in the  fact that the first order action
is the WZW action on the {\it complex} group, hence it  allows the
Wakimoto free
field representation (cf. \cite{FG}).  Of course, the symplectic form
in the
cotangent bundle representation can also be written in Darboux coordinates
 \cite{Ts},
but the Hamiltonian is not Gaussian. In the loop group $LSL(2,\bc)$
representation, the
whole action (not only the symplectic form part) can be written in
the Gaussian
way.

\section{ Poisson-Lie T-duality}
\noindent 1. The important structural feature of the $\ti G$-Poisson-Lie
symmetric $\si$-models
on $G$
 is their generic T-dualizability, i.e. (if a   condition 4   below is
 fulfilled
then) there is the dual $\si$-model on $\ti G$ which is
   moreover $G$-Poisson-Lie symmetric. In general,
  T-duality is a dynamical equivalence between
two (or more) $\si$-models living on targets with different geometry
 and/or topology.
The    Poisson-Lie T-duality is the particular form of the T-duality,
in which the common dynamics of all equivalent $\si$-models is given
by the
first order action (6) living on the Drinfeld double $D$.   Why the
action (6)
can encode more than one $\si$-model (9)? The point is simple, set
$\E_{\ti g}=Ad_{\ti g^{-1}}\E Ad_{\ti g}$ and suppose that
the conditions 1)-3) on page 4 are supplemented by another one:
\vskip1pc
 \noindent 4) the  $\ti G$-dependent operator $\ti\J\E_{\ti g}\J$ is
invertible.
\vskip1pc
\noindent
Then we may insert into the action (6) the dual counterpart of the
decomposition
(7):
$$l(\si,\tau)=\ti g(\si,\tau) h(\si,\tau),\qquad\ti g\in \ti G,
\quad  h\in  G.$$
Mimicking the derivation after Eq. (7), we eliminate the quantity
$\d_\si hh^{-1}$
and obtain the $\si$-model living on $\ti G$:
$$\ti S=\jp\int
 (\ti g^{-1}\d_+ \ti g,
(\ti\J \E_{\ti g}\J)^{-1}(\ti\J+\ti\J \E_{\ti g}\ti\J)
\ti g^{-1}\d_- \ti g)_\D.\eqno(33)$$
\vskip1pc
\noindent 2.
The duality between the $\si$-models (9) and (33) is the original
 Poisson-Lie
T-duality described in \cite{KS1}.  It was later generalized in
\cite{KS3} by noting,
that one can sometimes extract from  the first order action (6)
also $\si$-models
whose  targets are not Lie group manifolds. Indeed, suppose that
 there is
a subgroup $M$ of the double $D$, whose Lie algebra $\M$ is maximally
isotropic
(it  has thus the same dimension as $\G$ or $\ti\G$).   The group $D$
is then called
the Manin double\footnote{ Note a subtlety: we do not require the
 existence
of the complementary dual group $\ti M$ whose Lie algebra $\ti\M$ would be
 maximally
isotropic and $\D$ could be represented as the direct sum of the vector
spaces
$\M$ and $\ti\M$. If such a complementary group existed we would say that
 $D$
is the Drinfeld double of $M$.}   of $M$ \cite{AK}.
Consider then the right coset $D/M$ and parametrize\footnote{If there
exists no global
section of this fibration, we can choose several local sections covering
the whole
base space  $D/M$.} it by the
elements $f$ of $D$. With this parametrization of $D/M$, we may parametrize
the surface  $l(\tau,\si)$ in the double as follows
$$l(\tau,\si)=f(\tau,\si)m(\tau,\si),\quad  m\in M.\eqno(34)$$
 The action (6) then becomes
$$S=\jp\int (f^{-1}\d_\tau f,f^{-1}\d_\si f)_\D +
{1\over 12}\int d^{-1}(dff^{-1},[dff^{-1},dff^{-1}])_\D+$$
$$+\int (\d_\si mm^{-1},f^{-1}\d_\tau f)_\D-
\jp\int(f^{-1}\d_\si f  +\d_\si mm^{-1},\E_f(f^{-1}\d_\si f +
\d_\si mm^{-1}))_\D,\eqno(35)$$
where $\E_f=Ad_{f^{-1}}\E Ad_f$   and we tacitly suppose the measure
$d\si d\tau$ present in the formula. Now we note that the  expression (35)
is Gaussian in the $\M$-valued variable
$ \d_\si mm^{-1}$. In order to solve it away, we do not use a
 projector on $\M$ (an analogue of $\J$ or $\ti\J$) since
there is no canonically given kernel. The most useful strategy is to pick
up some basis $S^a$ in $\M$, write $\d_\si mm^{-1}=\mu_aS^a$ and integrate
away $\mu_a$. This gives
$$S=\js\int d\si d\tau ( \d_+ff^{-1}, \d_-ff^{-1})_\D+
{1\over 12}\int d^{-1}(dff^{-1},[dff^{-1},dff^{-1}])_\D+$$
$$+\jp\int d\si d\tau (f^{-1}\d_+ f,S^a-\E_f S^a)_\D(A_f^{-1})_{ab}
(S^b,f^{-1}\d_-f)_\D,\eqno(36)$$
where
$$A_f^{ab}=(S^a,\E_f S^b)_\D.\eqno(36b)$$
We thus conclude that  a $\si$-model (36) can be
associated to every maximally isotropic subalgebra of $\D$ provided the
matrix $A_f$ is invertible. The target of this $\si$-model is the coset
$D/M$. It turns out \cite{KS3} that the maximally isotropic subgroups $M$
and $M'$
 related by
a conjugation by a fixed element of the group $D$  give rise respectively
to the same
$\si$-models. The problem of classifying the Poisson-Lie dual (or plural)
models
thus reduces to the search of all maximally isotropic subalgebras of $\D$
up to conjugation \cite{KS3}.

We finish this section by giving the Euclidean analogue of the formula
(36),
taking place for $\E^2=-Id$:
$$S=-{i\over 4}\int d\si d\tau ( \d_zff^{-1}, \d_\bz ff^{-1})_\D+
{1\over 12}\int d^{-1}(dff^{-1},[dff^{-1},dff^{-1}])_\D+$$ $$
+\jp\int d\si d\tau (f^{-1}\d_z f,S^a+i\E_f S^a)_\D
(A_f^{-1})_{ab}(S^b,f^{-1}\d_\bz
f)_\D.\eqno(37)$$
In spite of the explicit presence of the imaginary unit in this formula,
the action (37) is always real.

\subsection{T-duality and the anisotropic model}

The Yang-Baxter $\si$-model (22) on the compact group $K=SU(2)$ is
left invariant
and right  Poisson-Lie symmetric. The left invariance can be also
interpreted \cite{KS1}
as the (left) Poisson-Lie symmetry corresponding to a trivial (null)
Poisson bracket
on $K$. The dual group    is then  the dual space $\K^*$ of $\K=Lie(K)$
and the group law of $\K^*$
is just the addition of vectors. The Lie algebra of the dual group
$\K^*$ is thus clearly Abelian.

So the Yang-Baxter $\si$-model  can be  dualized with respect to the
left $\K^*$-Poisson-Lie symmetry.
  The  dual
  $\si$-model has the well-known action \cite{OQ,KS1}:
$$S= \jp\int d\si d\tau \d_+\chi^a M_{ab}(\chi)\d_-\chi^b.\eqno(38)$$
Here  $\chi^a $ are the coordinates on the target $\K^*$
with respect to some chosen basis $t^*_a\in\K^*$. The matrix $M(\chi)$
is the inverse of
$$(M^{-1})^{ab}=(\ti R)^{ab}+f^{ab}_{~~~c}\chi^c,\eqno(39)$$
where $f^{ab}_{~~~c}$ are the structure constants of $\K=Lie(K)$ in the
basis $T^a$ dual to $t^*_a$ and $(\ti R)^{ab}=(T^a,\ti RT^b)_\D$.
We note that the duality between (22) and (38) is also referred
to as the traditional non-Abelian T-duality \cite{OQ}.

The dualization of the Yang-Baxter $\si$-model with respect to the right
 Poisson-Lie
symmetry gives the model (33). Note, however, that the operator $\E$,
though being
$K$-invariant, need not be $AN$-invariant.  In other words, the
dual model need not be Yang-Baxter
from the point of view of the dual group $\ti G$.  In any case, the action
 of the dual model
can be rewritten in the following nice way \cite{KS2}:
$$S=\jp\int d\si d\tau
 ( \d_+ bb^{-1},(\ti R+\ti\Pi_b)^{-1}\d_-bb^{-1})_\D.\eqno(40)$$
Here $b(\tau,\si)$ is the map from the worldsheet into the group
$\ti G=AN$ and
the operator $  \ti\Pi_b:\K\to\ti\G$  is given by
 $$\ti\Pi_b=\tJ Ad_b \J Ad_{b^{-1}}\J.\eqno(41)$$
We realize (cf. (13) and (14)) that $\ti \Pi_b$ defines the
Poisson-Lie bracket on $\ti G$.
It turns out also that the dual $\si$-model is right $G$-Poisson-Lie
symmetric
with respect to the right action of $\ti G$ on itself \cite{KS1,KS2}.

We thus observe that the Yang-Baxter $\si$-model (22) living on $K$
has two different
duals (38) and (40) living respectively on the groups $\K^*$ and
$\ti G=AN$.
 We may call this phenomenon the enhanced Poisson-Lie $T$-duality.
The dual $\si$-models are either of the form (40), if we dualize
(22) with respect to the right $AN$-Poisson-Lie
symmetry   (the Drinfeld double is $K^\bc$), or of the form (38)
if we dualize with respect to the left $\K^*$-Poisson-Lie
symmetry or, in other words, with respect
to the traditional non-Abelian duality (the Drinfeld double
is the cotangent bundle $T^*K$).

We give for completeness also the explicit description of the
dual model (40) living on $AN$.
The matrices $\ti R_\al$ in (40)
are given by (21)  and the Poisson-Lie operator
(41) is given by
$$\ti\Pi_b T^1= u_2t_3- [u_1^2+u_2^2 +\js e^{4\phi}-
\js  ]t_2;$$
$$\ti\Pi_b  T^2=- u_1t_3+  [u_1^2+u_2^2 +\js e^{4\phi} -
\js  ]t_1;$$
$$\ti\Pi_b T^3=  u_1t_2 - u_2t_1.$$
 Finally, in the parametrization (28), we have
$$(dbb^{-1})^1=e^{2\phi}d(e^{-2\phi}u_1),\quad (dbb^{-1})^2=
e^{2\phi}d(e^{-2\phi}u_2),
\quad (dbb^{-1})^3=d\phi.$$

\subsection{T-duality and the SL(2,\bc)/SU(2) WZW model}

Since the $\scu$ model has the Yang-Baxter property, it can be dualized
either with respect to the ordinary left symmetry or with respect to the
right Poisson-Lie symmetry. In the former case the Drinfeld double is the
cotangent bundle $T^*AN$ of the group $AN$, which has, up to conjugacy,
seven
maximally isotropic subalgebras, and in the latter it is $SL(2,\bc )$
with three maximally isotropic subalgebras.  One such subalgebra of
 $Lie(T^*AN)$ and one of
$sl(2,\bc)$ give the same $\scu$ theory. Thus it remains eight potentially
 new models.

From  the eight  cases we must remove  one isotropic subalgebra
 ($Lie(AN)$) of
$SL(2,\bc)$ and  one of $T^*AN$ because the respective matrices
 $A_f$ are
 not invertible (cf (36b)).
 Thus there remains
six  $\si$-models, but one of them is  not conformal due to tracefulness
of its
structure constants (cf.\cite{VU} for general discussion of this issue)
and other  three of them can be obtained also by Abelian duality with
spectators so we shall
not discuss them in any detail.
Thus there  remains a couple of possibly interesting  $\si$-models,
dual in a truly non-Abelain way to  the $\scu$ model. One of them is
based on the
double
$T^*AN$ and the other on $SL(2,\bc)$.
Here are the details:
\subsubsection{The double $T^*AN$}

The group law of $AN$ can be obtained from the matrix multiplication
in the parametrization (28).  Replacing $\phi$ by $\lm$ and $u_{1,2}$
by $L_{1,2}$,
the law reads
$$(L_1,L_2,\lm)(K_1,K_2,\k)=(L_1+e^{2\lm}K_1,L_2+e^{2\lm}K_2,\lm+\k).$$
The group law on the cotangent bundle $T^*AN$ is the standard one: the
semidirect
product of $AN$ acting in the coadjoint way
on   the dual of $Lie(AN)$ viewed as the Abelian additive group.
The resulting group is six-dimensional, it has the topology ${\br}^6$,
 and, in the
suitable parametrization, the group law turns out to be
$$ (L,\lm,l,w)(K,\k, k,y)=(L+e^{2\lm}K,\lm +\k,
 l+e^{-2\lm}k,w+y+2(L,k)e^{-2\lm}),$$
where $L=(L^1,L^2)$, $K=(K^1,K^2)$ and $(K,L)=K^1L^1+K^2L^2$.
We have
$$  M^{-1}\equiv (L,l,w,\lm)^{-1}=
(-e^{-2\lm}L,-\lm,-e^{2\lm}l,-w+2(L,l))$$
and the unit element is $e_d=(0,0,0,0)$.
\vskip1pc

\noindent The set of the left invariant vector fields:
The right  multiplication by $(\de L^\al,0,0,0)$ induces
$$ \nabla^R_{t_\al}=e^{2\lm}  {\d\over \d L^{\al}};$$
by $(0,0,\de l_\al,0)$ gives
$$ \nabla^R_{\tau^\al}=e^{-2\lm}\big({\d\over \d l_{\al}}+2L^{\al}
{\d\over \d w}\big);$$
and by $(0,0,0,\de w)$ and $(0,\de\lm,0,0)$ yields respectively
$$ \nabla^R_{\tau^0}={\d\over\d w},\quad \nabla^R_{t_0}={\d\over \d\lm}.$$
\noindent Hence we obtain the following bracket on $Lie(T^*AN)$:
$$ [t_0,t_\al]=2t_\al, \quad [t_\al,t_\beta]=0,\quad [\tau^i,\tau^j]=0,
\qquad i,j=0,1,2;$$
$$   [t_i,\tau^0]=0,\quad [t_0,\tau^\al]=-2\tau^\al,
\quad [t_\al,\tau^\beta]=
2\delta_\al^\beta \tau^0, \quad\al,\beta=1,2.$$
There is a non-degenerate  symmetric invariant bilinear form
on $Lie(T^*AN)$ defined by
$$ (\tau^i,\tau^j)_d=(t_i,t_j)_d=0,\qquad (\tau^i,t_j)_d=\de^i_j.
\eqno(42)$$
It is this form that makes $T^*AN$ the Drinfeld double of $AN$.

Now we would like to write down the duality invariant first order action
(6) on the double $T^*AN$.  Such action leads to $\si$-models which
are {\it right} symmetric or Poisson-Lie symmetric. Since we wish to
dualize
with respect to the {\it left} symmetry of our $\scu$ model
  we have to generalize our formalism in order to be
able to take into account also the left   Poisson-Lie
symmetric models. This generalization is very easy: if $L_R(g)$
is the Lagrangian of a right Poisson-Lie symmetric $\si$-model
then $L_L(g)=L_R(g^{-1})$ is   left  Poisson-Lie symmetric.

Thus we can consider the action (6) for the double $T^*AN$ and
we choose  the following operator $\E$:
$$\E t_1=-t_2,\quad \E t_2=t_1,\quad \E t_0 =2 \tau^0;$$
$$\E \tau^1=-\jp t_1 +\tau^2,\quad \E \tau^2=-\jp t_2 -\tau^1,
\quad \E \tau^0= -\jp t_0.$$
One easily checks that $\E$ is selfadjoint with respect to the form
(42) and
it holds $\E^2=-Id$.

The maximally isotropic subalgebras of $Lie(T^*AN)$ have been classified
(up to conjugacy) in \cite{DR}. There are two one-parametric families of
them
and five singular cases. Here is the list
$$1\gamma) \quad Span(\tau^0,\gamma_1\tau^1+\gamma_2\tau^2,
-\gamma_2 t_1+\gamma_1 t_2),\quad
\gamma_1^2+\gamma_2^2=1;$$
$$2\delta) \qquad Span(t_0, \delta_1 t_1+\delta_2 t_2,
-\delta_2 \tau^1+\delta_1 \tau^2),\quad
\delta_1^2+\delta_2^2=1;$$
$$3)\quad Span(t_0,\tau^2+t_1,\tau^1-t_2);\quad
4)\quad Span(t_0,\tau^2-t_1,\tau^1+t_2);$$
$$5)\quad Span(\tau^0,\tau^1,\tau^2);\quad
6) \quad Span(\tau^0,t_1,t_2)\quad 7) \quad
Span(t_0,t_1,t_2).$$
First of all, the choice of the subalgebra 5) leads to the
$\si$-model (37) living on the target $T^*AN/Lie^*(AN)=AN$.
Performing on it the transformation
$f\to f^{-1}$ leading from the right to the left symmetric model,
 we obtain precisely
the action (30) of the $\scu$ model. The choices 7) leads to a
 nonconformal
$\si$-model since the structure constants are tracefull. The
 possibility 6)  must
be also rejected since
the matrix (36b) is not invertible in this case.
The choices $ 1\gamma)$, 3)
and 4) lead to $\si$-models which can be also obtained by using the
standard
Abelian T-duality  with respect to the isometries of the $u,\bar u$
plane
(cf. (30)). We do not detail the corresponding $\si$-model metrics
and $H$-fields
here.

Thus the only truly non-Abelian case is $2\delta)$. It turns out that
all choices of $\delta$ gives the same $\si$-model, hence we take
$\delta_1=1,
\delta_2=0$. The corresponding   subgroup will be denoted as
$C=\exp{(Span(t_0,t_1,\tau^2))}$. The $\si$-model target is then
$T^*AN/C$
and can be identified with the group
$\ti C=\exp{(x_0\tau^0+x_1\tau^1+x_2 t_2)}$,
where $x_j$ are coordinates on the   group manifold $\ti C$.
Note that $C$ is the Poincar\'e group in two dimension and $\ti C$
is Abelian.
Inserting a decomposition $l=\ti cc$  ($c\in C,\ti c\in\ti C$)
in the first order action (6) and eliminating the variable $c$ gives
the following $\si$-model action:
$$S=\jp\int d\si d\tau\d_z x_k N^{kl}(x)\d_\bz x_l,$$
where
$$(N^{-1})_{kl}=
\left(\matrix{-2&2ix_1&-2ix_2\cr -2ix_1&0&1\cr 2ix_2&1&\jp}\right).$$
In particular, the metric part of the $\si$-model background is
$$ds^2={-dx_0^2 -(1+4x_2^2)dx_1^2-4x_1^2 dx_2^2 +(2-4x_1x_2)dx_1 dx_2
\over 2-2x_1^2-8x_1x_2}.$$
This metric changes signature (in some domain of the space-time
it is  Riemannian in another pseudo-Riemannian).
It is the matter of taste to say that it is either its very interesting
or rather pathological feature. We adopt the latter (and more coservative)
viewpoint and turn to the investigation of the double $SL(2,\bc)$.

\subsection{ The double $SL(2,\bc)$: $dS_3/AdS_3$ duality}  There are
 (up to conjugacy)
three  maximally isotropic subalgebras of $sl(2,\bc)$: $su(2)$,
$sl(2,\br)$ and
$Lie(AN)$. For $Lie(AN)$, the matrix $A_f$ (cf. (36b)) is not invertible,
hence it remains the   T-duality between two $\si$-models living
respectively
on the cosets $SL(2,\bc)/SU(2)=AN$ and $SL(2,\bc)/SL(2,\br)$. The former
 model
is just the $\scu$ model with the action (30), describing strings
in the Euclidean $AdS_3$ target and we shall see soon that the
$SL(2,\bc)/SL(2,\br)$
model
captures the dynamics of strings in the three-dimensional de Sitter
space $dS_3$. We recall that here the Poisson-Lie T-duality  is Euclidean,
 or,
 in other words,
it relates two $\si$-models with {\it real} Euclidean action. This means
 that
the de Sitter $H$-field is imaginary.

 The action of the $SL(2,\bc)/SL(2,\br)$ model is of course
given by the general formula (37). To make it more explicit we have to
specify  the
sections $f$ of the fibration $SL(2,\bc)/SL(2,R)$.  This requires some
preliminary exposition of the geometry of   this coset:

For convenience, we first    embed $SL(2,\br)$  into $SL(2,\bc)$ in  a
nonstandard way
$$\left(\matrix{\mu & i\nu\cr i\rho &\lm}\right)\in SL(2,C),\quad
 \mu,\nu,\rho,\lm\in\br,\quad \mu\lm +\nu\rho=1.$$
We shall  refer to the atypically embedded group $SL(2,\br))$ as
to $SL^a(2,\br)$.
We recall that $\si$-models (37) on $D/M$ and $D/M'$ have the same
target geometry
if $M$ and $M'$ are conjugated in $D$.
In fact, our embedding  $SL^a(2,\br)$ is conjugated
to the standard one (real matrices with unit determinant) by the following
element of
$SL(2,\bc)$  :
$${1\over  2}\left(\matrix{1+i & 1+i\cr i-1 &1-i}\right).$$
Consider  a space $dS_3$ of Hermitian 2$d$-matrices with determinant equal
 to  $(-1)$.
They can be written as
$$s=\left(\matrix{u&w\cr\bar w &v}\right), \quad uv-\bar ww=-1.\eqno(43)$$
Clearly, $dS_3$ is nothing but the de Sitter space, if it is equipped with
 the
Minkowski metric
$$ds_{dS}^2=dudv-d\bar wdw,\eqno(44)$$
restricted to the surface $uv-\bar ww=-1$. There exists a natural map
$s: SL(2,\bc)\to dS_3$ defined as
$$s(l)=l\si_1l^\dagger,\quad
\si_1=\left(\matrix{0&1\cr 1&0}\right).$$
It is clear that the preimage of the point $\si_1\in dS_3$ under the
map $s$
is a group, in fact, it is our group $SL^a(2,{\bf R})$.
Thus we see that the map $s$ defines also an injective map from
the coset  $SL(2,\bc)/SL^a(2,R)$ to the de Sitter space $dS_3$. By a small
abuse
of notation we call this map also $s$. It is not difficult to prove
that the map $s:  SL(2,\bc)/SL^a(2,R)\to dS_3$ is also surjective.

Our next goal is to find the section $f$ of the fibration
$SL(2,\bc)/SL^a(2,R)$
to be inserted in the action (6). To do this,
we consider first  the Iwasawa decomposition of $SL^a(2,R)$ :
$$\left(\matrix{\mu & i\nu\cr i\rho &\lm}\right)=
\left(\matrix{\cos{\th} & i\sin{\th}\cr i\sin{\th} &\cos{\th}}\right)
\left(\matrix{1 & iM  \cr 0 &1}\right)
\left(\matrix{e^\phi & 0\cr 0 &e^{-\phi}}\right),
 \phi,M\in{\bf R}, 0\leq \th\leq 2\pi,$$ and the  Iwasawa decomposition
of  $SL(2,\bc)$:
$$ \left(\matrix{a & c \cr b& d}\right)=
\left(\matrix{\al & -\bar\bt\cr \bt &\bar\al}\right)
\left(\matrix{1 & L \cr 0 &1}\right)
\left(\matrix{1 & iM  \cr 0 &1}\right)
\left(\matrix{e^\phi & 0\cr 0 &e^{-\phi}}\right), \eqno(45)$$
where $a,b,c,d\in \bc$, $\phi,L,M\in\br$ and $\al,\bt\in\bc$
with $\al\bar\al+\bt\bar\bt=1$.
We recognize in the first matrix on the r.h.s. of (45) an   element
of the group $SU(2)$.
Comparing those two Iwasawa decompositions, we arrive
at the conclusion, that we can choose $f$ within the form
$\left(\matrix{\al & -\bar\bt\cr \bt &\bar\al}\right)
\left(\matrix{1 & L \cr 0 &1}\right)$.
We remark the following identity
$$\left(\matrix{\cos{\th}+i{L \over \sqrt{L^2+1}} \sin{\th} &
 i{1\over \od}\sin{\th}
 \cr  i{1\over \od}\sin{\th} &\cos{\th}-i{L \over \sqrt{L^2+1}}
\sin{\th}}\right)\left(\matrix{1 & L \cr 0 &1}\right)=$$
$$=\left(\matrix{1 & L \cr 0 &1}\right)\left(\matrix{\cos{\th} &
 i\od\sin{\th}\cr i{1\over\od}\sin{\th}
 &\cos{\th}}\right).\eqno(46)$$
The second matrix on the r.h.s. of (46) is clearly in $SL^a(2,\br)$,
 hence the
product of two matrices on the l.h.s. is in the same coset as
 $\left(\matrix{1 & L \cr 0 &1}\right)$. Moreover, the first matrix
on the l.h.s. of (46)
is in $SU(2)$. We call it $R_L(\th)$ and we note that
$$R_L(\th_1+\th_2)=R_L(\th_1)R_L(\th_2).$$
For $\th$ close to zero it holds
$$R_L(\th)=\left(\matrix{1 & 0 \cr 0 &1}\right)+i{\th\over \od}
 \left(\matrix{L &  1\cr 1 &-L}\right)+O(\th^2)\eqno(47)$$
and it is true also that
$$Tr\biggl\{\left(\matrix{L &  1\cr 1 &-L }\right)
\left(\matrix{0 & 1\cr -1 &0}\right)\biggr\}=0.\eqno(48)$$
The facts (47) and (48) imply that there is an $L$-dependent global
Gaussian decomposition
of $SU(2)$:
$$\left(\matrix{\al & -\bar\bt\cr \bt &\bar\al}\right)=
\left(\matrix{\cos{\vth}+
i{L \over \sqrt{L^2+1}} \sin{\vth} & i{1\over \od}\sin{\vth}
 \cr i{1\over \od}\sin{\vth} &\cos{\vth}
-i{L \over \sqrt{L^2+1}}\sin{\vth}} \right)\times$$
$$\times
\left(\matrix{\cos{\chi} & \sin{\chi}\cr -\sin{\chi} & \cos{\chi}}\right)
 \left(\matrix{\cos{\th}+i{L \over \sqrt{L^2+1}} \sin{\th} &
 i{1\over \od}\sin{\th}
 \cr   i{1\over \od}\sin{\th} &\cos{\th}
-i{L \over \sqrt{L^2+1}}\sin{\th}} \right),$$
where $0\leq \vth,\th\leq \pi$ and $0\leq \chi\leq \pi/2$.
Regarding the identity (46), we find immediately the needed section $f$:
$$f=\left(\matrix{\cos{\vth}+
i{L \over \sqrt{L^2+1}} \sin{\vth} & i{1\over \od}\sin{\vth}
 \cr i{1\over \od}\sin{\vth} &\cos{\vth}
-i{L \over \sqrt{L^2+1}}\sin{\vth}} \right)
\left(\matrix{\cos{\chi} & \sin{\chi}\cr -\sin{\chi} &
 \cos{\chi}}\right)\left(\matrix{1 & L \cr 0 &1}\right),\eqno(49)$$
where $0\leq \vth\leq \pi$, $0\leq \chi\leq \pi/2$ and  $L\in\br$.

\vskip1Pc
\noindent The map $s(f)$ now gives the transformation between
the "cylindrical"
coordinates $\vth,\chi,L$ and the de Sitter parametrization (43):
$$\jp(u+v)=L,\quad \jp(u-v)=L\cos{2\chi}+\sin{2\chi}\cos{2\vth},
\eqno(50a)$$
$$w=\cos{2\chi}-L\sin{2\chi}
\cos{2\vth}-i\od\sin{2\chi\sin{2\vth}}.\eqno(50b)$$
The reader may verify that, indeed, it holds $\quad uv-\bar ww=-1$.
\vskip1Pc
\noindent
Now we read off from (37)   the metric and the three-form field $H$ of
the $\si$-model
on $SL(2,{\bf C})/SL^a(2,\br)$:
 $$ds^2= (f^{-1}df,S^a)_\D(A_f^{-1})_{ab}(S^b,f^{-1}df)_\D;$$
$$H=+{i\over 2}d\{(f^{-1}df,\E_fS^a)_\D\w (A_f^{-1})_{ab}
 (S^b, f^{-1}df)_\D\}-i
{1\over 12} (dff^{-1}\st [dff^{-1}\st dff^{-1}])_\D.$$
We choose the following basis of the $Lie(SL^a(2,\br))$:
$$S^+=\left(\matrix{ 0&i\cr 0&0}\right),\quad S^-=
\left(\matrix{ 0&0\cr i&0}\right),
\quad S^0=\jp\left(\matrix{ 1&0\cr 0&-1}\right).$$
Recall that the operator $\E=\I$ is $SL(2,\bc)$-invariant,
hence $\E_f=\I$
and the matrix
$A_f$ (cf. (36b)) is also independent on $f$. The non-zero components
of its inverse are
$$(A^{-1}_f)_{+-}=(A^{-1}_f)_{-+}=-1,\quad (A^{-1}_f)_{00}=2.$$
Considering (49), it is straightforward to calculate the form $f^{-1}df$
and
hence the metric and the $H$-field of the
$SL(2,{\bf C})/SL^a(2,\br)$ model:
$$ds^2=-\jp {(dL)^2\over L^2+1}\sin^2{2\chi}\sin^2{2\vth}+
2(L^2+1)(d\chi^2+\sin^2{2\chi}d\vth^2)+$$
$$+2\cos{2\vth}dL d\chi
-\sin{4\chi}\sin{2\vth}dLd\vth;\eqno(51a)$$
 $$H=-4i\od \sin{2\chi} dL\w d\chi\w d\vth .\eqno(51b)$$
By using (50ab), we can check easily
 that (up to a normalization factor $1/2$) the metric (51a) is nothing
but the standard de Sitter metric (44).

Because the Lie algebra $sl(2,\br)$ does not have the traceful
structure constants
(i.e. $f^{ab}_{~~b}= 0$),
we expect to find a dilaton $\Phi$ such that the background (51ab)
will satisfy the following
conditions \cite{CI,KT} of the conformal invariance
$$R_{mn}-\js H_{mpq}H_n^{~pq}+2D_m D_n\Phi=0,
\quad -\jp D_nH^n_{~pq}+\d_n\Phi H^n_{pq}=0.
\eqno(52)$$
Here $R_{mn}$  means the Ricci tensor and  $D_n$ the covariant derivative.
 The computation indeed shows that the conditions (52) are satisfied
for $\Phi=0$.
 \subsection{T-duality and loop group symmetry}
It is well-known that the $\scu$ model has  a $SL(2,\bc)$ current symmetry.
 It
can be very transparently
understood in the first order formalism, where the action
  reads (cf. (24))
 $$S(l)=\jp\int d\si d\tau {\rm ImTr}(\d_\bz ll^{-1}\d_\si ll^{-1})
+{1\over 12}\int d^{-1}{\rm ImTr}(dll^{-1}\w [dll^{-1}\st dll^{-1}]).
\eqno(24)$$
By using the Polyakov-Wiegmann formula \cite{PW}
$$S(l_1l_2)=S(l_1) +S(l_2)+
\int  d\si d\tau {\rm ImTr}(l_1^{-1}\d_\bz l_1 \d_\si l_2l_2^{-1}),
\eqno(53)$$
the field equations of the first order model (24)  can now be easily
obtained.
We set in (53) $l_1=l$ and $l_2=1+\chi(z,\bz)$, with
$\chi\in sl(2,{\bf C})$
being infinitesimal and we  find
$$\de S=\int  d\si d\tau {\rm ImTr}(l^{-1}\d_\bz l \d_\si \chi).$$
Since $\d_\si\chi$ is arbitrary and the form ${\rm ImTr}$ non-degenerate,
we conclude that the first order field equations have extremely
simple form:
$$l^{-1}\d_\bz l=0.\eqno(54)$$
 The symmetries of the first order equation (54) are evident: a solution
$l(z)$ becomes clearly another solution upon the transformation
$$l(z)\to g(z)l(z), \quad g(z)\in \slc.\eqno(55)$$
Our next goal is to visualize the huge symmetry (55) in the
  second order   $\scu$ model and its $SL(2,{\bf C})/SL^a(2,\br)$
de Sitter dual.
\vskip1Pc
\noindent 1. $SL(2,{\bf C})/SU(2)$ model. Recall that the
target space is the group $AN$ and the Lagrangian is given by (30).
We can perform the change of variables on the target $AN$ as follows
 $$h=bb^\dagger=\left(\matrix{ e^{2\phi}+4e^{-2\phi}u\bar u&
2e^{-2\phi}\bar u\cr
2e^{-2\phi}u&e^{-2\phi}}\right).$$
In terms of the matrix $h$, the $\si$-model background (31) can
 be rewritten
as
$$ds^2_{AdS}={1\over 8}{\rm Tr} (dhh^{-1} dh h^{-1}),\eqno(56a)$$
$$H={1\over 48}{\rm Tr}(dhh^{-1}\w [dhh^{-1}\st dhh^{-1}]).\eqno(56b)$$
 In other words, the $\scu$ action (30) can be rewritten
as
$$S=i\int dz\w d\bz {\rm Tr}(\d_z hh^{-1}\d_\bz hh^{-1})
+{i\over 6}\int d^{-1}{\rm Tr}(dhh^{-1}\w[hh^{-1}\st dhh^{-1}]).\eqno(57)$$
 Note that in (56ab) and (57) there is the full trace, not only its
imaginary
part.

\vskip1Pc
\noindent How the loop symmetry (55)  manifests itself from the point
of view of the second order action (57)?  First note that from the
first order trajectory $l$ we obtain the second order trajectory
$b$ by the Iwasawa decomposition $l=bk$, $b\in AN$, $k\in SU(2)$
(cf. (25)).
This means that
$$h=ll^\dagger$$
and the symmetry (55) becomes
$$h(z,\bz)\to g(z)h(z,\bz) g^\dagger(\bz).\eqno(58)$$
Indeed, by the direct use of the corresponding Polyakov-Wiegmann formula, 
we see
that the model (57) has the symmetry (58).

\vskip1Pc
\noindent 1. $SL(2,{\bf C})/SL^a(2,\br)$ model. Recall that the
target is the space $dS_3$ of Hermitian matrices with determinant (-1):
$$s=\left(\matrix{u&w\cr \bar w&v}\right),\quad uv-w\bar w=-1.$$
Using  the parametrization (50) and ignoring an unimportant
overall normalization,  the $\si$-model metric and $H$-field
 (51ab) can be rewritten as
$$ds^2_{dS}={\rm Tr} (dss^{-1} dss^{-1}),$$
$$H={1\over 6}{\rm Tr}(dss^{-1}\w [dss^{-1}\st dss^{-1}]).$$
 In other words, the de Sitter $\si$-model  action  can be written
as
$$S=i\int dz\w  d\bz {\rm Tr} (\d_z ss^{-1}\d_\bz ss^{-1})
+{i\over 6}\int d^{-1}{\rm Tr}(dss^{-1}\w[dss^{-1}\st dss^{-1}]) .
\eqno(59)$$
 Note that  the only distiction  of the $AdS_3$ action  (57)
and its $dS_3$ dual (59)
is the sign of the respective determinants of the Hermitian matrices
$h$ and $s$.

\vskip1Pc
\noindent How the loop symmetry (55)  manifests itself from the point
of view of the second order action (59)?  First note that from the
first order trajectory $l$ we obtain the second order trajectory
$f$ by the decomposition   $l=fm$,   $m\in SL^a(2,\br)$ (cf. (34)).
This means that
$$s=f\si_1f^\dagger=l\si_1l^\dagger.$$
and the symmetry (55) becomes
$$s(z,\bz)\to g(z)s(z,\bz) g^\dagger(\bz).\eqno(60)$$
Indeed, by the direct use of the Polyakov-Wiegmann formula, we see
that the model (59) has the symmetry (60).
\section{Conclusions and outlook}
We have shown that the simultaneous presence of the ordinary (left)
symmetry
and the Poisson-Lie (right) symmetry of $\si$-models on group targets
results
in a rich $T$-duality picture. Our main example of a unitary theory
was the anisotropic principal chiral model (22) and its duals (38) and
(40).
 However, it seems that
a better candidate for settling the quantum status of Poisson-Lie
T-duality
would be   the dual pair of $AdS_3$ and $dS_3$ models (57) and (59).
 There
is a surprising feature of this duality, namely, the $dS_3$ background has
a different signature than the (Euclidean)
$AdS_3$ one.  Further investigations are needed
to elucidate this fact.

In the subsequent paper, we plan to describe how the $dS_3/AdS_3$
T-duality
works for open strings and $D$-branes.


\begin{thebibliography}{19}
\bibitem{OQ}{X. de la Ossa and F. Quevedo, {\it Nucl. Phys.} {\bf B403}
(1993) 377; B.E. Fridling and A. Jevicki, {\it Phys. Lett.} {\bf B134}
(1984) 70;
E.S. Fradkin and A.A. Tseytlin, {\it Ann. Phys.} {\bf 162} (1985) 31}
\bibitem{KS1}{C. Klim\v c\'\i k and P. \v Severa,
{\it Phys. Lett.} {\bf B351}
(1995) 455; C. Klim\v c\'\i k, {\it Nucl. Phys. (Proc. Suppl.)} {\bf
B46} (1996) 116; P. \v Severa,
{\it Minim\'alne plochy a dualita}, Diploma thesis, 1995, in Slovak}

\bibitem{Ch}{I. V. Cherednik; Theor. Math. Phys. 47 (1981) 225}
 \bibitem{KS2}{C. Klim\v c\'\i k and P. \v Severa,
 {\it Phys. Lett.} {\bf B372} (1996) 65}
\bibitem{D}{ V.G. Drinfeld, {\it Quantum groups}, in
{\it Proceedings ICM}, Berkeley (1986) 708;
 A. Yu. Alekseev and A.Z. Malkin,
{\it Commun. Math. Phys.} {\bf 162} (1994) 147}
\bibitem{FG}{F. Falceto and K. Gaw\c edzki, {\it J. Geom. Phys.}
{\bf 11} (1993) 251}
\bibitem{Gaw}{K. Gaw\c edzki, {\it Non-compact WZW conformal field
theories}, in: Proceedings
of NATO ASI Carg\`ese 1991, eds. J. Fr\"ohlich, G. t'Hooft, A. Jaffe,
G. Mack, P.K. Mitter,
R. Stora. Plenum Press (1992) 247}
\bibitem{Ts}{J. Teschner, {\it Nucl. Phys.} {\bf B546} (1999) 390}

\bibitem{KS3}{C. Klim\v c\'\i k and P. \v Severa,
 {\it Phys. Lett.} {\bf B383} (1996) 281}
\bibitem{AK}{A. Alekseev and Y. Kosmann-Schwarzbach, {\it Manin pairs
and moment maps},
math.DG/9909176}
\bibitem{VU}{A. Giveon and M. Ro\v cek, {\it Nucl. Phys.} {\bf B421}
(1994) 173;
E. \'Alvarez, L. \'Alvarez-Gaum\'e and Y. Lozano, {\it Nucl. Phys.}
{\bf B424} (1994) 155;
R. von Unge, {\it JHEP} {\bf 0207} (2002) 14}
\bibitem{DR}{S. Deragopian  and A. Roulland, {\it Classification des
 sous-alg\`ebres
isotropes
de $Lie(T^*AN)$}, Diploma thesis, University Aix-Marseille II, 2002,
in French}
\bibitem{CI}{ C. G. Callan, D. Friedan, E. Martinec and M. J. Perry,
{\it Nucl. Phys.} {\bf B262} (1985) 593;
E.S. Fradkin and A.A. Tseytlin, {\it Nucl. Phys.} {\bf B261} (1985) 1;
A. A. Tseytlin, Phys. Lett. {\bf B178} (1986) 349}
\bibitem{KT}  {C. Klim\v c\'\i k and A. A. Tseytlin, {\it Nucl. Phys.}
{\bf B424} (1994) 71}
\bibitem{PW}{A.Polyakov and P.B.Wiegmann, {\it Phys. Lett.}
{\bf B311} (1983) 549}







\end{thebibliography}
\end{document}